\documentstyle[preprint,aps,epsf]{revtex}
\newcommand{\Sla}[1]{{\not \! \! {#1}}}
\newcommand{\sla}[1]{{\not \! \! \: {#1}}}
\newcommand{\slaD}{\sla{D}}  

\newcommand{\slad}{\Sla{\, d}}
\begin{document}

\title{Axial Anomaly Effect in Chiral p-wave Superconductor}

\author{
Jun Goryo\footnote{present address; Inst. Solid State Phys., 
U. Tokyo, Kashiwanoha, Kashiwa 277-0882, Japan.  
e-mail; jfg@issp.u-tokyo.ac.jp}} 

\address
{Yukawa Institute for Theoretical Physics, Kyoto University, Kyoto, 
606-8502 Japan.}

\maketitle

\begin{abstract}
We analyze the chiral $p$-wave superconductor in the low temperature
region. The superconductor has a 
$\varepsilon_{x} p_{x}+i \varepsilon_{y} p_{y}$-wave gap   
in two-dimensional space (2D). 
Near the second superconducting transition point, 
the system could be described by a quasi-1D  
chiral $p$-wave model in 2D.  
The axial anomaly occurs in such a model and causes an accumulation of 
the quasiparticle in an inhomogeneous magnetic field. 
The effect is related to the winding number of the gap.
\end{abstract}

 
\newpage

Chiral $p$-wave superfluidity is realized 
in the superfluid $^{3}$He-A \cite{volovik-1}. 
Recently, the possibility of 
chiral $p$-wave superconductivity has been argued 
in Sr$_{2}$RuO$_{4}$\cite{cpsc}. 
In such superfluids or superconductors,  
the condensation of the Cooper pairs, which have orbital angular 
momentum along the same direction, occurs. 
Therefore, time-reversal symmetry (T) and 
also parity (P) in two-dimensional space (2D) are violated. 
We investigate an axial anomaly effect in the chiral $p$-wave superconductor 
near the second superconducting transition point\cite{s-j-r}.    
It is revealed that the axial anomaly causes 
P- and T-violating phenomena related to the topologically quantized number 
at sufficiently low temperatures.   

The axial anomaly was originally pointed out 
in the Dirac QED in 3D\cite{a-b1,a-b2,b-j}. It is the phenomenon 
in which a symmetry under the phase transformation $e^{i \gamma_{5} \alpha}$ 
of the Dirac field in the action 
at the classical level is broken in the quantum theory. 
Here, $\alpha$ is a constant and $\gamma_{5}$ is a hermitian 
matrix which anti-commutes with all of the Dirac matrices $\gamma_{\mu}$, 
where $\mu$ is the spacetime index. 
The Adler-Bardeen's theorem guarantees the absence of higher order 
corrections to the divergence of the axial current\cite{a-b1,a-b2}. 
Therefore, the exact calculation of the two-photon decay rate 
of the neutral $\pi$ meson can be carried out.
It has been pointed out that the same results are 
obtained by using the path-integral formalism and  
the relation between the axial anomaly and topological 
quantized numbers has been clarified through the Atiyah-Singer index
theorem\cite{fujikawa1,fujikawa2,index}. 

It has been pointed out that the axial anomaly also plays an important role  
in the quantum Hall effect (QHE) in the 2D massive Dirac QED.   
In 2D, the mass term of the Dirac fermion violates P and T similar to  
the magnetic field, and 
the Hall effect may occur. 
It was shown that the existence of the Hall current and its quantization  
are caused by the axial anomaly in 1D\cite{nie-sem,red,ishikawa1,ishikawa2}.
       
The relation between the axial anomaly and QHE     
in 2D electron gas in the magnetic field was 
also discussed, 
and the quantized Hall conductance is 
expressed by the winding number of the fermionic propagator in 
the momentum space\cite{ishikawa1,ishikawa2,i-i-m-t1,i-i-m-t2,i-i-m-t3}.  
Other applications of the axial anomaly 
to condensed matter physics has been studied in 
the field of the superfluid $^{3}$He in 3D
\cite{volovik-1,he-3-anomaly1,he-3-anomaly2},   
and also in some 1D systems
\cite{cdw-anomaly,nag-oshi}. 

The phenomena caused by the axial anomaly are  
related to the topologically quantized numbers.  
On the analogy of QHE, it is expected that the axial anomaly 
also plays an important role in other P- and T-violating 2D systems. 
In this letter, we show an axial anomaly effect in a chiral $p$-wave
superconductor in 2D, which has the spin-triplet 
$\varepsilon_{x} p_{x} + i \varepsilon_{y} p_{y}$-wave symmetry. 
P and T-violation occur whenever 
both $\varepsilon_{x}$ and $\varepsilon_{y}$ are nonzero. 
We show that near the second superconducting phase transition
(2ndSCPT) point the axial anomaly in 1D causes an accumulation of    
the mass density of the quasiparticle in an inhomogeneous magnetic field.  
The effect is related to the winding number of   
the gap; ${\rm sgn}(\varepsilon_{x} \varepsilon_{y})$\cite{vol,g-i}. 
%
We use the 2+1-dimensional Euclidian spacetime and 
the natural unit ($\hbar=c=1$) in the present paper.

Let us consider 2ndSCPT in the chiral $p$-wave state. 
It has been proposed theoretically that such a transition 
occurs under uniaxial pressure in the $x-y$ plane (the 
basal plane)\cite{s-j-r}. 
2ndSCPT caused by an exact in-plane 
magnetic field has also been argued\cite{agterberg}, but 
our argument is not applicable to this transition because 
the anomaly effect which we will show requires a magnetic field
perpendicular to the basal plane.  
The orbital dependence of the gap {\it near} the transition point is shown in 
Fig. 1. For simplicity, 
we assume a circular fermi surface in the normal state.  
There are two tiny gap points around ${\bf p}=(\pm p_{\rm F},0)$,
where $p_{\rm F}$ is the Fermi momentum. 
Therefore, the quasiparticle excitations along the $x$-axis 
become dominant in the low temperature region. 
We can describe such a situation effectively by using 
a quasi-1D model. For the present, we consider the {\it superfluid} 
just for simplicity, and will extend our argument to 
the supercondcuting case by taking into account the Meissner effect. 
We assume a linearized fermion spectrum and 
a spin-triplet chiral $p$-wave gap near the Fermi surface 
in the normal state (i.e. $p_{x}=\pm p_{\rm F} + \delta p_{x}$ ,
$\delta p_{x} << p_{\rm F}$) written as,  
\begin{eqnarray}
\epsilon_{\rm R,L}({\bf p})&=&\pm v_{\rm F} \delta p_{x} , 
\label{udkin}\\ 
\Delta({\bf p}) 
&=& i \sigma_{3} \sigma_{2} 
\frac{\Delta}{p_{\rm F}}(\varepsilon_{x} p_{x} +  i \varepsilon_{y} p_{y})
\nonumber\\
&\simeq& i \sigma_{3} \sigma_{2} 
\Delta(\pm \varepsilon_{x} +  i \varepsilon_{y} 
\frac{p_{y}}{p_{\rm F}}), 
\label{c-p-w}
\end{eqnarray} 
where 
$v_{\rm F}$ is the Fermi velocity. 
$\epsilon_{R} ({\bf p})$ ($\epsilon_{L} ({\bf p})$) is 
the kinetic energy for the right (left) mover. 
$\varepsilon_{x}$ and $\varepsilon_{y}$ are parameters and presently   
they satisfy $\varepsilon_{x}<<1$ and $\varepsilon_{y}\sim 1$.

The Lagrangian of our model is written as 
\begin{eqnarray} 
{\cal {L}}&=&\bar{\Psi}_{\bf p} 
\left[\{ i \partial_{\tau} + \mu \frac{d {\rm B}_{z}}{d y} y \sigma_{3} \} 
\otimes \gamma_{\tau} \right.   
\label{ucpsc-lag}\\
&&\left.+ v_{\rm F} \delta p_{x} \hat{1} \otimes \gamma_{x} 
+ \frac{\Delta}{p_{\rm F}} 
\varepsilon_{y} p_{y} \sigma_{3} \otimes \gamma_{y}
-i \Delta  \varepsilon_{x} \sigma_{3} \otimes \hat{1}\right] \Psi_{\bf p}.
\nonumber
\end{eqnarray} 
Here, we use the Bogoliubov-Nambu representation with an isospin $\alpha=1,2,$
$$
\Psi({\bf x})=e^{i p_{\rm F} x} 
\left(\begin{array}{c}
\psi_{\rm R}({\bf x})  \\ 
i \sigma_{2} \psi^{*}_{\rm L}({\bf x})  
\end{array} \right),  
$$ 
and $\Psi_{\bf p}$ is its Fourier transform. 
$\psi_{\rm R}({\bf x})$ and $\psi_{\rm L}({\bf x})$ are
the slowly varying fields for the right mover and the left mover 
with a real spin index $s=1,2$, 
respectively.  The matrices 
$\gamma_{\tau}=\tau_{1}, \gamma_{x}=\tau_{2}$ and $\gamma_{y}=\tau_{3}$ 
are the $2\times2$ Pauli matrices with isospin indices and 
$\sigma_{i}~(i=1,2,3)$ are the $2\times2$ Pauli matrices with spin indices.  
The symbol $\sigma_{i} \otimes \gamma_{\tau,x,y}$ represents 
the direct product.
$\bar{\Psi}$ is defined as $\bar{\Psi}= - i \Psi^{\dagger} \gamma_{\tau}$.   
We assume a magnetic field, which is directed to the $z$-axis 
(the $c$-axis in the crystal) and  
has a constant gradient in the $y$-direction, 
i.e., $B_{z}(y)=(d B_{z} / d y) y$, and $(d B_{z} / d y)=const.$ 
Because we are considering the superfluid for the present, 
the magnetic field couples with the fermion only through the Zeeman term 
$\mu B_{z} \bar{\Psi} \sigma_{3} \otimes \gamma_{\tau} \Psi$, 
where $\mu$ is the magnetic moment of the fermion. 
In general, the magnitude of the gap $\Delta$ depends on the magnetic field, 
but we neglect this fact because it has become apparent that our result  
does not depend on $\Delta$. 

We note that the Lagrangian is equivalent to that of the 
2D Dirac QED\cite{fisher} in a constant electric field, except for 
the existence of $\sigma_{3}$.   
It has been shown that the axial anomaly causes the quantized Hall 
effect in the 2D Dirac system\cite{nie-sem,red,ishikawa1,ishikawa2}.  
In our model, the physical value which corresponds to the quantized
Hall current 
$j_{x}=e \left< \bar{\Psi}_{\rm D} \gamma_{x} \Psi_{\rm D} \right>$  
in 2D Dirac theory ($\Psi_{\rm D}$ is the Dirac field and 
$<\cdot\cdot\cdot>$ is the expectation value for the ground state) is the 
mass (charge) density, because $e \left<\bar{\Psi} \gamma_{x} \Psi \right> = 
e \left< \psi^{\dagger}_{\rm R} \psi_{\rm R} + 
\psi^{\dagger}_{\rm L} \psi_{\rm L} \right> = e \left<\rho_{e} \right>$,   
where $e$ represents a mass of the quasiparticle,     
and represents an electric charge when we consider a superconductor.
The reason why $e \left<\bar{\Psi} \gamma_{x} \Psi \right>$ becomes the
charge density, not the current, is the remarkable difference between 
$\Psi_{\rm D}$ and $\Psi$. $\Psi_{\rm D}$ is a two-component fermion 
field and each component has the same 
charge. $\Psi$ is also two component (two isospin) but the
signs of their charges are opposite.  

Let us calculate the expectation value of the mass (charge) density  
\begin{equation}
\left<\rho_{e}({\bf x})\right> = {\rm Tr}\left[
\frac{e\sigma_{3} \gamma_{x}} 
{i \slad \gamma_{x} + v_{\rm F} \delta p_{x}\sigma_{3} \otimes \gamma_{x}  
- i \Delta \varepsilon_{x}} \right]. 
\label{j-exp}\\
\end{equation}
A hermitian operator $\slad$ in 1D (in the $y$-direction) is defined as 
\begin{equation} 
\slad = \left\{ i \partial_{\tau} \sigma_{3} +  
\mu \frac{d B_{z}}{d y} y \hat{1}\right\} \otimes \tilde{\gamma}_{\tau} 
\nonumber\\
+ 
\frac{\Delta}{p_{\rm F}} \varepsilon_{y} p_{y} 
\hat{1} \otimes \tilde{\gamma}_{y},    
\end{equation} 
where $\tilde{\gamma}_{\tau} = - \gamma_{y} = - \tau_{3}$ and 
$\tilde{\gamma}_{y} = \gamma_{\tau} = \tau_{1}$. 
We define $\gamma_{5}$ as 
$\gamma_{5}=i \tilde{\gamma}_{\tau} \tilde{\gamma}_{y}=-\gamma_{x}=-\tau_{2}$,
and it is anti-commute with $\tilde{\gamma}_{\tau}$ and $\tilde{\gamma}_{y}$, 
therefore, $\gamma_{5}$ is a hermitian matrix which satisfies 
$\{ \gamma_{5}, \slad \}= 0$. 
These findings suggest that 
if an eigenstate $u_{n}$ of $\slad$ with a nonzero 
eigenvalue $\xi_{n}$ ($0<n$) exists 
(i.e. $\slad u_{n} = \xi_{n} u_{n}$), 
$\gamma_{5} u_{n}$ should be another eigenstate with an eigenvalue $-\xi_{n}$.
If zero modes of $\slad$ exist (i.e. $\slad u_{0} = 0$ and 
$\slad \gamma_{5} u_{0}=0$), they are divided into two groups. 
One of them is $u_{0}^{(+)}=(1/2)(1+\gamma_{5})u_{0}$ 
with an eigenvalue $\gamma_{5}=+1$ and 
the other is $u_{0}^{(-)}=(1/2)(1-\gamma_{5})u_{0}$ with 
an eigenvalue $\gamma_{5}=-1$, since $\gamma_{5}^{2}=1$.

Let us research the eigenmodes of $\slad$. 
The expectation value of $\slad^{2}$ is 
\begin{eqnarray} 
(u_{n}, \slad^{2} u_{n})&=&|\omega_{c}| 
(n + \frac{1}{2}) 
+ \frac{\omega_{c}}{2} (u_{n}, \gamma_{5} u_{n}),   
\nonumber\\ 
\omega_{c}&=&\mu \frac{d B_{z}}{d y} \frac{2 \Delta}{p_{\rm F}} 
\varepsilon_{y}, 
\label{expvalue}
\end{eqnarray}   
where $u_{n}=u_{n}(y - y_{c}(p_{\tau}, \sigma_{3}))$ 
is the eigenfunction of 
the harmonic oscillator with the frequency  
$\omega_{c}$. The oscillator is centered at 
$
y_{c} (p_{\tau},\sigma_{3}) = 
- (d B_{z} / d y)^{-1} (p_{\tau}/\mu) \sigma_{3}. 
$
Equation (\ref{expvalue}) indicates that 
only zero modes which belong to $u_{0}^{-}$ ($u_{0}^{+}$) exist 
when $0<\omega_{c}$ ($\omega_{c}<0$).    
This suggests the non-conservation of the vacuum expectation value of 
the axial charge which is defined in the second-quantized formalism as 
\begin{eqnarray}
\left<Q_{5}\right>&=&\left<N_{+} - N_{-}\right>,  
\nonumber\\ 
N_{\pm}&=&\int dp_{y} \hat{u}_{0}^{\dagger (\pm)} \hat{u}_{0}^{(\pm)},  
\end{eqnarray} 
while the classical 1D theory  
${\cal{L}}_{\rm 1D}=\bar{\Psi} \slad \Psi$  
has the axial symmetry $\Psi \rightarrow e^{i \alpha \gamma_{5}} \Psi$,   
i.e., {\it the axial anomaly occurs.} 
Here, $\hat{u}_{0}^{\pm}$ is a second-quantized fermionic field. 
The anomaly comes from the spectral asymmetry 
of zero modes, as in the discussions presented  
in Refs.\cite{nie-sem,red,ishikawa1,ishikawa2,j-l-h}.  
In the free system, the energy spectrum of $u_{0}^{(\pm)}$ is 
$p_{0}= \pm \frac{\Delta}{p_{\rm F}} \varepsilon_{y} p_{y}$ 
for spin down ($\sigma_{3}=+1$) states in Minkowski spacetime, 
and all of the negative energy states are filled   
while all of the positive energy states are empty and $\left<Q_{5}\right>=0$. 
After we turn on the magnetic field adiabatically 
(for the present, we assume $0<\omega_{c}$),  
the energy spectrum of $u_{0}^{(+)}$ is lowered   
and $\left<N_{+}\right>$ decreases (i.e., empty negative energy states arise 
on the spectrum of $u_{0}^{(+)}$); on the other hand,  
the energy spectrum of $u_{0}^{(-)}$ is lifted   
and $\left<N_{-}\right>$ increases (i.e., filled positive energy states arise 
on the spectrum of $u_{0}^{(-)}$), 
and therefore, $\left<{Q_{5}}\right>$ does {\it not} conserve.  
Finally $\left<N_{+}\right>=0$ and only $u_{0}^{(-)}$ exists.  
We see that the same argument is also valid for spin down ($\sigma_{3}=-1$)
states. 
The nonzero eigenvalues of $\slad^{2}$ are  
$E_{n}=\omega_{c} (n + 1/2)$, since 
the inner product $(u_{n}, \gamma_{5} u_{n})$ vanishes whenever 
$\slad u_{n} \neq 0$ because of the orthogonal relation between  
the eigenfunctions of the hermitian operator.

Next, we consider the eigenvalue problem of a 2D operator: 
\begin{equation}
\slaD = \slad -  
\Delta \varepsilon_{x} \hat{1} \otimes \gamma_{x} 
= \slad +  
\Delta \varepsilon_{x} \hat{1} \otimes \gamma_{5}.
\end{equation}
Let 
\begin{equation}
\varphi_{n}= (\alpha_{n} u_{n} + \beta_{n} \gamma_{5} u_{n}) 
e^{i p_{x} x}
\end{equation}
stand for an eigenfunction. 
We use a representation for the $n$-th level such as 
\begin{equation} 
\slad= 
\left( \begin{array}{cc} 
\xi_{n} & 0  \\
0 & - \xi_{n}  
\end{array} \right) , 
u_{n}=
\left( \begin{array}{c} 
1 \\
0
\end{array} \right) ,  
\gamma_{5}=  
\left( \begin{array}{cc} 
0 & 1 \\
1 & 0 
\end{array} \right), 
\end{equation} 
where, 
$$
\xi_{n}=
\left\{\begin{array}{cl}
\sqrt{|\omega_{c}| (n + \frac{1}{2})} 
& (n = 1,2,\cdot\cdot\cdot), 
\\
0 & (n=0). 
\end{array}\right.
$$
Therefore, the eigenvalue equation is written as  
\begin{equation}
\left(\begin{array}{cc}
\xi_{n} & \Delta \varepsilon_{x}  \\ 
\Delta \varepsilon_{x} & -\xi_{n}  
\end{array} \right)
\left(\begin{array}{c}
\alpha_{n} \\ 
\beta_{n} 
\end{array} \right)
=\zeta_{n}
\left(\begin{array}{c}
\alpha_{n} \\ 
\beta_{n} 
\end{array} \right).   
\end{equation} 
There are two eigenstates for an oscillator in the $n (\neq 0)$-th level  
written as  
\begin{eqnarray}
&\zeta_{n}^{(\pm)}(p_{x})&
=\pm \sqrt{\xi_{n}^{2} + 
\Delta^{2} \varepsilon_{x}^{2}},  
\\
&\left(\begin{array}{c}
\alpha_{n}^{+} \\
\beta_{n}^{+} 
\end{array} \right)&  
=\frac{1}{C_{+}} 
\left(\begin{array}{c} 
\zeta_{n}^{(+)} + \xi_{n}  \\
\Delta \varepsilon_{x}  
\end{array} \right) ,  
\nonumber\\
&\left(\begin{array}{c}
\alpha_{n}^{-} \\
\beta_{n}^{-} 
\end{array} \right)&
=\frac{1}{C_{-}} 
\left(\begin{array}{c} 
- \Delta \varepsilon_{x} \\ 
-\zeta_{n}^{(-)} + \xi_{n} 
\end{array} \right), 
\nonumber
\end{eqnarray}
where $C_{\pm}$ are normalization constants, but for $n=0$, there is only 
one eigenstate   
\begin{eqnarray} 
\zeta_{0}(p_{x})&=& \frac{\omega_{c}}{|\omega_{c}|} 
\Delta |\varepsilon_{x}|   
,  
\nonumber\\
\left(\begin{array}{c} 
\alpha_{0} \\
\beta_{0} 
\end{array} \right) 
&=& \frac{1}{2}
\left(\begin{array}{c} 
1 \\
- \omega_{c}/|\omega_{c}| 
\end{array} \right), 
\end{eqnarray} 
because the solution should satisfy 
$\gamma_{5} \varphi_{0} = - (\omega_{c}/|\omega_{c}|)\varphi_{0}$. 
This condition comes from the axial anomaly in the $y$-direction.  

Finally, we show the accumulation of 
the mass density from Eq. (\ref{j-exp}),  
which is derived as
\begin{eqnarray} 
\left<\rho_{e}({\bf x})\right>&=& {\rm Tr} 
\left[ 
\frac{-i e \sigma_{3}} 
{\slaD - i v_{\rm F} \delta p_{x} \sigma_{3} \otimes \hat{1}} 
\right] 
\nonumber\\ 
&=& \sum_{n} 
\int_{-\infty}^{\infty} \frac{d p_{\tau}}{2 \pi} 
\int_{-\Lambda}^{\Lambda} \frac{d \delta p_{x}}{2 \pi} {\rm tr} \left[ 
\frac{- i e \sigma_{3}   
|u_{n}(y - y_{0}(p_{\tau}, \sigma_{3}))|^{2}}  
{\zeta_{n}(p_{x})  - 
i v_{\rm F} \delta p_{x} \sigma_{3}} \right] 
\nonumber\\
&=& \frac{e \mu}{2 \pi}\frac{d B_{z}}{d y} \sum_{n \neq 0}    
\int_{-\Lambda}^{\Lambda} \frac{d \delta p_{x}}{2 \pi} {\rm tr} \left[ 
\frac{1}  
{\zeta_{n}^{(+)}(p_{x})  - 
i v_{\rm F} \delta p_{x} \sigma_{3}} 
+
\frac{1}  
{\zeta_{n}^{(-)}(p_{x})  - 
i v_{\rm F} \delta p_{x} \sigma_{3}} 
\right]
\nonumber\\
&&
+  \frac{e \mu}{2 \pi} \frac{d B_{z}}{d y}    
\int_{-\Lambda}^{\Lambda} \frac{d \delta p_{x}}{2 \pi} {\rm tr} \left[ 
\frac{1}  
{\zeta_{0}(p_{x}) - 
i v_{\rm F} \delta p_{x} \sigma_{3}} \right]
\nonumber\\
&=& {\rm sgn}(\varepsilon_{x} \varepsilon_{y}) 
e \mu N_{\rm 1D}(0) 
\frac{d B_{z}}{d y},    
\label{charge-density}
\end{eqnarray}
where the symbol $tr$ represents a trace on the real spin, 
and we use the normal-orthogonal relation $\int dy |u_{n}|^{2}=1$.     
$\Lambda$ is a momentum cutoff and   
we assume a relation $|\Delta|<<\Lambda^{2} / 2 m<<\epsilon_{F}$.    
$N_{\rm 1D}(0)=(2 \pi v_{\rm F})^{-1}$ is the density of state at 
the Fermi surface in 1D.
All of the $n \neq 0$ parts are canceled out because of the 
co-existence of the eigenvalues $\zeta_{n}^{(+)}$ and $\zeta_{n}^{(-)}$.     
{\it Only the $n=0$ part survives because of the axial anomaly in the 
$y$-direction}. 

The effect is related to the winding number of   
the gap in Eq. (\ref{c-p-w})\cite{vol,g-i},   
\begin{eqnarray}
\int \frac{d^{2} p}{16 \pi} 
tr[\hat{{\bf g}} \cdot 
({\bf \nabla}\hat{{\bf g}} \times {\bf \nabla} \hat{{\bf g}})]  
&=&{\rm sgn}(\varepsilon_{x} \varepsilon_{y})
, 
\\  
{\bf g}({\bf {p}})&=& 
\left(\begin{array}{c} 
{\rm Re}[\Delta({\bf {p}})(-i \sigma_{2})]  \\ 
-{\rm Im}[\Delta({\bf {p}})(-i \sigma_{2})] \\
({\bf p}^{2} / 2 m) - \epsilon_{\rm F}
\end{array} \right),  
\nonumber
\end{eqnarray}
where ${\bf \nabla}=\partial / \partial {\bf p}$. 
This suggests that these effects occur even if $\varepsilon_{x}$ and/or  
$\varepsilon_{y}$ are infinitesimally small, and that these effects 
come from the P- and T-violation of the gap.

The accumulated mass density exists in the bulk region of the superfluid.  
In the superconductors, the Meissner effect occurs 
and the magnetic field cannot penetrate into the bulk, therefore  
the accumulated charge density would exist near the edge\cite{note}
of the superconductors and also around the vortex.

Let us roughly estimate the charge distribution around the vortex under
uniaxial pressure. 
We make an approximation that the magnetic field is linearly decreased 
around 
the vortex core and vanishes at $r=\lambda_{\rm L}$, where $r$ is
the distance from the core and $\lambda_{\rm L}$ is the
London penetration depth; therefore, the gradient of the magnetic field 
is $-H_{c2} / \lambda_{\rm L}$, where $H_{c2}$ is an upper critical field 
along the $c$-axis. We use the parameters appropriate for 
Sr$_{2}$RuO$_{4}$, 
$H_{c2}=0.075$ Tesla, $\lambda_{\rm L}=1800$ $\AA$ and 
$v_{\rm F}=10^{6}$ $cm/s$. 
The induced charge density around the 
vortex core obtained from Eq.(\ref{charge-density}) 
is of the order of $-10^{-8}e$ per $1$ $\AA^{2}$, when 
${\rm sgn} (\varepsilon_{x} \varepsilon_{y})=+1$. 
For the charge neutrality, 
there should be a plus charge $\pi$ $\times$ (1800 [$\AA$])$^{2}$ $\times$
10$^{-8}e$ $[\AA ^{-2}]=10^{-1} e$ at the core.  
The core charge might extend in the region $r < \xi$ for the
stability. 
Here $\xi=\pi v_{\rm F} / \Delta$ is the coherent length. 
Further (self-consistent) calculation is required to obtain more 
accurate results. 

In summary, we have analyzed the 2D chiral $p$-wave superconductor 
near the second superconducting transition point\cite{cpsc,s-j-r} at 
sufficiently low temperature. Such a situation is described by a
quasi-1D model in 2D, which can be mapped onto the 2D Dirac QED. 
Corresponding to the QHE in 2D Dirac
theory\cite{nie-sem,red,ishikawa1,ishikawa2}, we show that the charge
accumulation occurs under an inhomogeneous magnetic field as an effect 
of the axial anomaly in 1D. The effect is related to the winding
number of the gap\cite{vol,g-i}, and could contribute to the vortex 
core charge.

Recently, the vortex in chiral superconductors has 
been  discussed\cite{goryo}, and such a vortex has a fractional charge and 
a fractional angular momentum.   
Interesting phenomena related to these fractional quantum numbers 
and the present effects are expected to occur 
around the vortex core.   

An anisotropic chiral $p$-wave state, which has the  
${\bf d}({\bf k})= \hat{\bf z} (\sin k_{x} + i \sin k_{y})$ order 
parameter, is proposed\cite{miyake-narikiyo}. 
This state has four tiny gap points, therefore our argument would be
applicable to quasiparticle excitations in the low temperature 
region of such a state\cite{goryo-et-al}.

The author is grateful to K. Ishikawa, R. Joynt, N. Maeda, K. Shizuya 
and M. Sigrist for useful discussions and encouragement.

\begin{figure}
\centerline{\epsfysize=4cm\epsffile{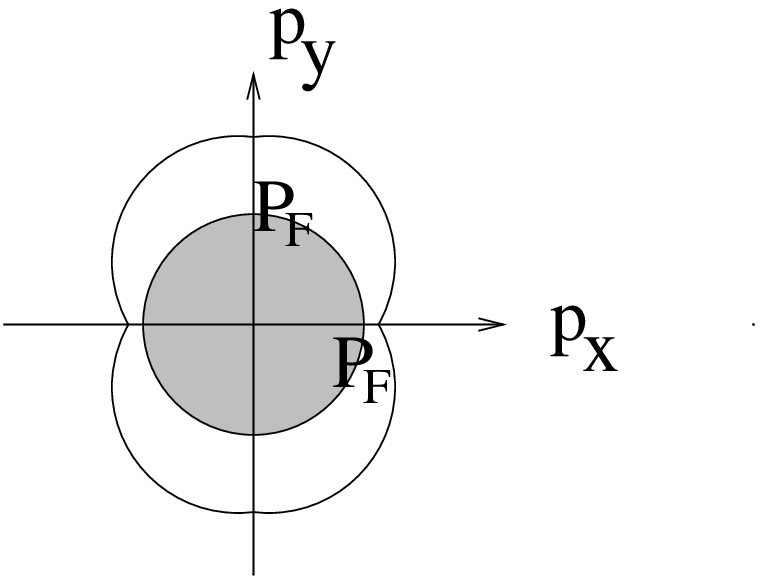}}
\caption{The Fermi sea in the normal state and 
the momentum dependence of the gap functions 
for Sr$_{2}$RuO$_{4}$ near the second superconducting 
phase transition point.    
The shadows show the Fermi sea, and the distance between 
outer lines and inner lines shows the magnitude of the gap $|\Delta /
p_{\rm F}|^{2} 
(\varepsilon_{x}^{2} p_{x}^{2} + \varepsilon_{y}^{2} p_{y}^{2})$. }
\label{fig:1}
\end{figure}

\end{document}